\documentclass[twocolumn,superscriptaddress,aps,preprintnumbers]{revtex4-1}

\usepackage{graphicx}
\usepackage{bm}
\usepackage{color}
\usepackage{epstopdf}
\usepackage{amsmath}
\usepackage{amssymb}
\usepackage{epstopdf}

\usepackage[urlcolor=blue,colorlinks=true,citecolor=blue,linkcolor=blue,pdfstartview={FitH},bookmarks=false]{hyperref}

\graphicspath{{fig/}{./fig/}{.}}

\begin{document}

\title{Bound states induced by the ferromagnetic dimer in a triangular lattice}

\author{Szczepan G\l{}odzik}
\email[e-mail: ]{szglodzik@kft.umcs.lublin.pl}
\affiliation{Institute of Physics, M.\ Curie-Sk\l{}odowska University, 
pl. M. Curie-Sk\l{}odowskiej 1, 20-031 Lublin, Poland}

\author{Andrzej Ptok}
\email[e-mail: ]{aptok@mmj.pl}
\affiliation{Institute of Nuclear Physics, Polish Academy of Sciences, 
ul. E. Radzikowskiego 152, 31-342 Krak\'{o}w, Poland}

\date{\today}

\begin{abstract}
The ongoing efforts aiming at control and manipulation of novel topological phases renewed the interest in bound states induced in superconducting substrates by magnetic impurities. 
First described by Yu, Shiba and Rusinov, those bound states are spin-polarized and exhibit a rather large spatial extent in two dimensional (2D) materials.
Here, using the Bogoliubov--de~Gennes formalism, we study the Yu--Shiba--Rusinov bound states induced by a ferromagnetic dimer in 2D triangular lattice. 
We describe various topographies of the in-gap bound states, depending on the coupling between impurities and superconducting host. 
\end{abstract}

\maketitle

\section{Introduction}
\label{sec.intro}

Local breaking of time-reversal symmetry -- stemming from the presence of magnetic impurities -- induces a pair of low energy bound states in superconducting hosts~\cite{balatsky.vekhter.06}.
Such states, first described by Yu, Shiba and Rusinov~\cite{yu.65,shiba.68,rusinov.69} are spin-polarized and the energy required to excite them changes depending on the strength of magnetic interaction with the bulk.
Recent progress in engineering topological superconductivity in nanowires~\cite{mourik.2012,lee.jiang.13,deng.vaitiekenas.16,nichele.drachmann.17} and impurities~\cite{nadj.perge.2014,ruby.peng.16,jeon.xie.17,cornils.kamlapure.17,island.gaudenzi.17} stimulated the increase in both theoretical and experimental studies concerning magnetic impurities (see e.g.~\cite{heinrich.pascual.17} for a review).
The pursuit to obtain stable, topologically non-trivial phases has extended from one-dimensional to two-dimensional lattices of the magnetic impurities~\cite{rontynen.ojanen.15,pershoguba.bjornson.15,li.2016,menard.guissart.16,rontynen.ojanen.16,weststrom.poyhonen.16,rachel.2017,bjornson.blackschaffer.17}, however to fully understand the phenomena in those rich systems, it is useful to study a simpler system, i.e. a dimer of impurities embedded in a superconductor.


Characteristic  star-like topography of YSR bound states can be realised e.g. in NbSe$_{2}$ crystal~\cite{menard.guissart.15}, when the impurity is treated clasically, i.e. the spin $S$ of the impurity is taken as infinite, and the interaction term $JS$ is kept finite and treated as a parameter~\cite{ptok.glodzik.17}. 
Experimentally the classical nature of the impurity can be probed by looking for a zero-bias peak in the conductance -- the manifestation of Kondo screening. 
When it is absent (cf. ref~\cite{menard.guissart.15}) the impurity is assumed to be of classical nature.
The aforementioned properties inspired experimental studies of YSR bound states induced by two impurities and the interplay between them. 
Experimental studies~\cite{kezilebieke.dvorak.17} concerning cobalt phthalocyanine (CoPC) molecules on NbSe$_{2}$, resolved the energy splitting of YSR bonding and anti-bonding states.

\begin{figure}[!b]
\centering
\includegraphics[width=0.85\linewidth]{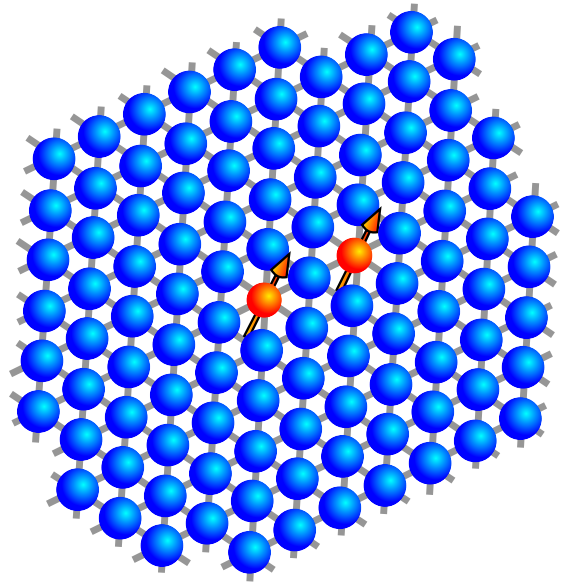}
\caption{
Schematical view of the studied system. 2D superconductor with triangular lattice (blue spheres) and two classical magnetic impurities, interpreted as magnetic moments localized at two lattice sites (orange spheres with arrows).
}
\label{fig.schem}
\end{figure}

In this paper, we focus on the topography of YSR bound states induced by ferromagnetic dimers in a superconductor with triangular lattice, presented schematically in Fig.~\ref{fig.schem}. 
Using the Bogolubov--de~Gennes (BdG) equations we obtain spatially resolved local density of states (LDOS) and examine the change in polarisation of the system, which originates from the quantum phase transition (QPT) associated with change of the ground state of the system, manifested by the ``crossing'' of bound states.

The paper is organized as follows: in Section~\ref{sec.model} we present the microscopic model and discuss some methodological details.
Next, in Section~\ref{sec.num} we show the numercial results and discuss them.
Finally, in Section~\ref{sec.sum} we summarize our results.

\section{Model and method}
\label{sec.model}

We describe the magnetic impurities embedded in a 2D superconducting host by the following Hamiltonian $\hat{\mathcal{H}} = \hat{\mathcal{H}}_{0} + \hat{\mathcal{H}}_{imp} + \hat{\mathcal{H}}_{int}$. 
The single particle term:
\begin{eqnarray}
\hat{\mathcal{H}}_{0} = - t \sum_{ \langle i,j \rangle \sigma } \hat{c}_{i\sigma}^{\dagger} \hat{c}_{j\sigma} - \mu \sum_{i\sigma} \hat{c}_{i\sigma}^{\dagger} \hat{c}_{i\sigma},
\end{eqnarray}
describes the kinetic energy of electrons. 
Here $\hat{c}_{i\sigma}^{\dagger}$ ($\hat{c}_{i\sigma}$) denotes creation (annihilation) of electron with spin $\sigma$ at {\it i}-th lattice site,
$t$ is the hopping integral between the nearest-neighbors and $\mu$ is the chemical potential. 
Assuming a large spin $S$ of the impurities allows to treat their interaction classically~
\cite{balatsky.vekhter.06}, such that the interaction potential becomes the following
\begin{eqnarray}
\hat{\mathcal{H}}_{imp} =  \sum_{i\sigma} \left( K - \sigma J \right) \hat{c}_{0\sigma}^{\dagger} \hat{c}_{0\sigma}.
\end{eqnarray}
 $J$ and $K$ are the magnetic and non-magnetic parts of the impurity potential, respectively.

We model the superconducting state by the on-site interaction 
\begin{eqnarray}
\hat{\mathcal{H}}_{int} = U \sum_{i} \hat{c}_{i\uparrow}^{\dagger} \hat{c}_{i\uparrow} 
\hat{c}_{i\downarrow}^{\dagger} \hat{c}_{i\downarrow}
\end{eqnarray}
with attractive potential $U < 0$. 
Because we assume the existence of superconducting order in the system and different distribution of average number of spin-up and spin-down particles, resulting from the ferromagnetic nature of the dimer, we choose the following mean-field decoupling:
\begin{eqnarray}
\hat{c}_{i\uparrow}^{\dagger} \hat{c}_{i\uparrow} \hat{c}_{i\downarrow}^{\dagger} \hat{c}_{i\downarrow} &=& \chi_{i} \hat{c}_{i\uparrow}^{\dagger} \hat{c}_{i\downarrow}^{\dagger} + \chi_{i}^{\ast} \hat{c}_{i\downarrow} \hat{c}_{i\uparrow} - | \chi_{i} |^{2} \\
\nonumber &+& n_{i\uparrow} \hat{c}_{i\downarrow}^{\dagger} \hat{c}_{i\downarrow} + n_{i\downarrow} \hat{c}_{i\uparrow}^{\dagger} \hat{c}_{i\uparrow} - n_{i\uparrow} n_{i\downarrow} ,
\end{eqnarray}
where $\chi_{i} = \langle \hat{c}_{i\downarrow} \hat{c}_{i\uparrow} \rangle$ is the superconducting order parameter (SOP) and $n_{i\sigma} = \langle \hat{c}_{i\sigma}^{\dagger} \hat{c}_{i\sigma} \rangle$ is the average number of particles with spin $\sigma$ at {\it i}-th site.

The Hamiltonian $\hat{\mathcal{H}}$ can be diagonalized via the Bogoliubov-Valatin transformation:
\begin{eqnarray}
\label{eq.bvtransform} \hat{c}_{i\sigma} = \sum_{n} \left( u_{in\sigma} \hat{\gamma}_{n\sigma} - \sigma v_{in\sigma}^{\ast} \hat{\gamma}_{n\bar{\sigma}}^{\dagger} \right) 
\end{eqnarray}
where $\hat{\gamma}_{n}$ and  $\hat{\gamma}_{n}^{\dagger}$ are the {\it quasi}-particle fermionic operators, with the eigenvectors $u_{in\sigma}$ and $v_{in\sigma}$. 
This leads to the BdG equations~\cite{degennes.99}:
\begin{eqnarray}
\label{eq.bdg} \mathcal{E}_{n\sigma} 
\left(
\begin{array}{c}
u_{in\sigma} \\ 
v_{in\bar{\sigma}}
\end{array} 
\right) 
&=& \sum_{j} \left(
\begin{array}{cc}
H_{ij\sigma} & D_{ij} \\ 
D_{ij}^{\ast} & -H_{ij\bar{\sigma}}^{\ast} 
\end{array} 
\right) 
\left(
\begin{array}{c}
u_{jn\sigma} \\ 
v_{jn\bar{\sigma}} 
\end{array} 
\right)
\end{eqnarray}
containing the single-particle term $H_{ij\sigma} = - t \delta_{\langle i,j \rangle} - \left( \mu - U n_{i\bar{\sigma}} - ( K - \sigma J ) \delta_{i0} \right) \delta_{ij}$.
$D_{ij} =  \Delta_{i} \delta_{ij}$ describes the on-site pairing, where $\Delta_{i}=U\langle \hat{c}_{i\downarrow}\hat{c}_{i\uparrow}\rangle$ is the gap function.

The superconducting order parameter $\chi_{i}$ and occupancy $n_{i\sigma}$ have to be computed self-consistently from BdG equations~(\ref{eq.bdg}):
\begin{eqnarray}
\chi_{i} &=&  \langle \hat{c}_{i\downarrow} \hat{c}_{i\uparrow} \rangle \\ \nonumber  &=& 
\sum_{n} \left[ u_{in\downarrow} v_{in\uparrow}^{\ast} f( \mathcal{E}_{n} ) 
- u_{in\uparrow} v_{in\downarrow}^{\ast} f ( - \mathcal{E}_{n} ) \right] , \\
n_{i\sigma} &=&  \langle \hat{c}_{i\sigma}^{\dagger} \hat{c}_{i\sigma} \rangle \\ \nonumber &=& 
\sum_{n} \left[ | u_{in\sigma} |^{2} f( \mathcal{E}_{n} ) + 
| v_{in\bar{\sigma}} |^{2} f ( - \mathcal{E}_{n} ) \right] ,
\end{eqnarray}
where $ f ( \omega ) = 1 / \left[ 1 + \exp ( \omega / k_{B} T ) \right]$ 
is the Fermi-Dirac distribution.
This type of calculations have been previously successfully used in case of disordered systems~\cite{maska.mierzejewski.03,maska.sledz.07,krzyszczak.domanski.10,ptok.12,ptok.kapcia.15}.
Using the transformation~(\ref{eq.bvtransform}) we can find the spin dependent local density of states (LDOS)~\cite{matsui.sato.03}:
\begin{eqnarray}
\nonumber \rho_{i\sigma} ( \omega ) = \sum_{n} \left[ | u_{in\sigma} |^{2} \delta 
( \omega - \mathcal{E}_{n} ) + | v_{in\sigma} |^{2} \delta ( \omega + \mathcal{E}_{n} ) 
\right] . \\
\label{eq.ldos}
\end{eqnarray}
The total density of states (DOS) is then given as $\sum_{i\sigma} \rho_{i\sigma} ( \omega )$.
The Dirac delta function has been replaced
by a Lorentzian $\delta (\omega) = \zeta / [ \pi ( \omega^{2} 
+ \zeta^{2})]$ with broadening $\zeta / t = 0.025$.

\section{Numerical results and discussion}
\label{sec.num}

\begin{figure}[!b]
\centering
\includegraphics[width=\linewidth]{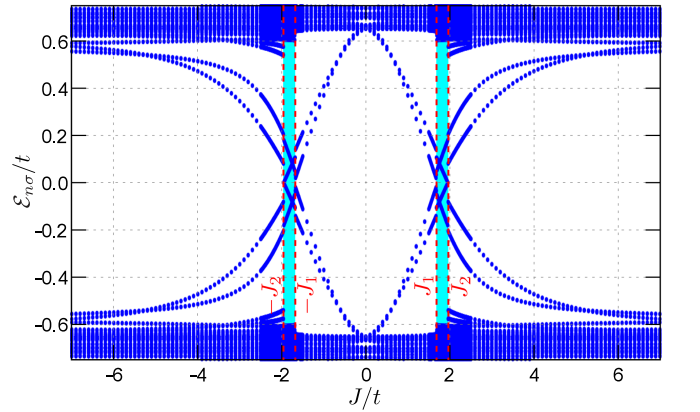}
\caption{
The eigenvalues of the studied system versus the magnetic interaction $J$. The light-blue area indicates the range of $J$ at which the polarisation $\mathcal{P}= 1/2 $.
}
\label{fig.warw}
\end{figure}

Numerical calculations were performed for the system comprising $N_{a} \times N_{b} = 41 \times 41$ sites, assuming $U/t = - 3$, $\mu/t = 0$ and $K/t = 0$.

Self-consistent solution of the BdG equations~(\ref{eq.bdg}) provides eigenvalues and eigenvectors of the Hamiltonian $\hat{\mathcal{H}}$ in the presence of disorder.
The spectrum of $\hat{\mathcal{H}}$ is shown in Fig.~\ref{fig.warw}.
We observe that the self-consistently achieved value of the superconducting energy gap is $U \langle \chi_{i} \rangle / t \simeq 0.6$. 
When the magnetic interaction $J$ increases, a pair of YSR bound states emerges from the gap edges and approaches the Fermi level.
At $J/t \simeq 0.8$ one can see that the hybridization of YSR states induced by the ferromagnetic dimer results in splitting into bonding and anti-bonding states. 
At some critical value $J = J_{1}$ the first pair of states crosses the Fermi energy, which induces a quantum phase transistion (QPT), associated with a change of the ground state \cite{salkola.balatsky.97}.
Additionally, the polarization of the system, defined as $\mathcal{P} = \sum_{i} ( n_{i\uparrow} - n_{i\downarrow} ) / 2$ changes from $0$ to $1/2$.
The fact that at $J = J_{2}$ the second pair of YSR bound states reaches the chemical potential implies that there should be an additional QPT.
Indeed, after the second crossing $\mathcal{P}$ increases again by $1/2$, reaching the value of $1$. In fact in a general case, number of the QPTs in the system depends on the number of the adatoms~\cite{meng.klinovaja.15,mohanta.kampf.17}. Each crossing at the Fermi level corresponds to reduction of the value of the order parameter. This is associated with the fact that when Shiba states cross at the Fermi energy, a state which is in phase with Cooper pairs of the bulk becomes unoccupied, while an out-of-phase state becomes occupied~\cite{bjornson.balatsky.17}.
Further increasing the strength of magnetic interaction leads to a growth of excitation energy of the bound states.
Additionally, both QPTs have the properties of a discontinuous phase transition~\cite{glodzik.ptok.17}.

\begin{figure}[!t]
\centering
\includegraphics[width=\linewidth]{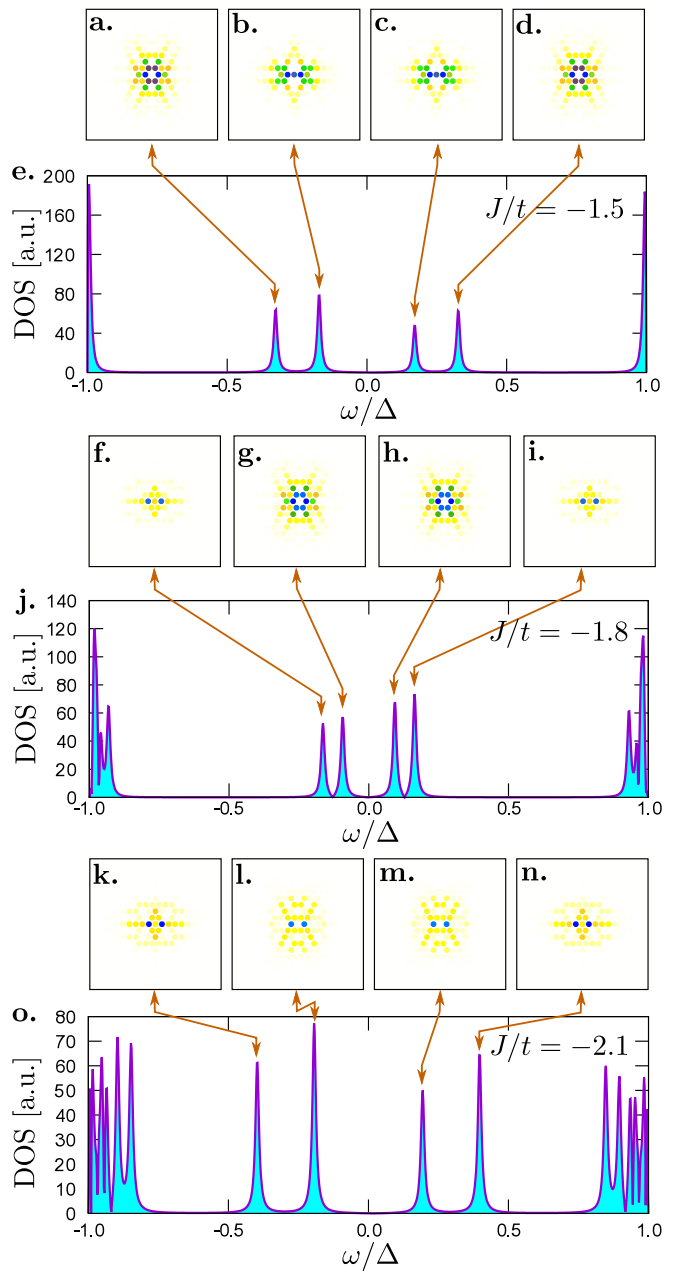}
\caption{
\
Subgap density of states for different values of magnetic coupling $J/t$: $-1.5$ (e), $-1.8$ (j) and $-2.1$ (o).
Panels a-d, f-i and k-n show localization of the Yu--Shiba--Rushinov bound states given by the $\xi_{in\uparrow}+\xi_{in\downarrow}$ (see Eq.~\ref{eq.xi}) corresponding to peaks marked by arrows.
}
\label{fig.dos}
\end{figure}

Now we will discuss the change of localization of the YSR bound states for different values of magnetic coupling $J$.
In order to do this, we define the quantity:
\begin{eqnarray}
\xi_{in\sigma} = | u_{in\sigma} |^{2} \theta ( - \mathcal{E}_{n} ) + | v_{in\sigma} |^{2} \theta ( \mathcal{E}_{n} ) ,
\label{eq.xi}
\end{eqnarray} 
which describes the localization of YSR bound states, with particle (hole) character in the case of $\mathcal{E}_{n} < 0$ ($\mathcal{E}_{n} > 0$). $\theta ( x )$ is the Heaviside step function.
Using $\xi_{in\sigma}$, the LDOS can be expressed as 
\begin{eqnarray}
\rho_{i\sigma} ( \omega ) = \sum_{n} \xi_{in\sigma} \left[ \delta 
( \omega - \mathcal{E}_{n} ) + \delta ( \omega + \mathcal{E}_{n} ) 
\right] .
\end{eqnarray}
In consequence $\xi_{in\sigma}$ can be treated as a partial component of the LDOS given by {\it n}-th eigenpairs of the BdG eigenproblem~(\ref{eq.bdg}).
Results of numerical calculations are shown in Fig.~\ref{fig.dos}.
Panels (e), (j) and (o) show the in-gap DOS for three different scenarios: 
before the first QPT, i.e. when $| J | < J_{1}$, after the first crossing of YSR states $J_{2} > | J | > J_{1}$ (cf. light-blue area in Fig.~\ref{fig.warw}), and for $ | J | > J_{2}$, respectively. 
In every case, there is a pair of split, asymmetric YSR bound states near the Fermi energy. 
For higher values of $J$ additional states emerge from the gap edges and in the case of $J/t = 2.1$, one of the pairs appears to be split due to the hybridization of the dimer.

Smaller panels above the DOS plots present the topography of bound states, given by $\xi_{in\uparrow}+\xi_{in\downarrow}$ (see Eq.~(\ref{eq.xi})). 
Central blue dots are the sites with the highest spectral value of bound states (i.e. impurity sites).
In the scenario with one impurity the $C_{6}$ symmetry of the triangular lattice is ''inherited'' by the bound states and the star-like shapes are obseved~\cite{menard.guissart.15,ptok.glodzik.17}.
Here, this symmetry is broken due to the interference of YSR bound states induced by separated impurities.
We observe that the states with the same absolute value of energy exhibit the same spatial profile and differ only in spectral weight. 
There is no significant change in the range of bound states after the QPTs, as for every value of $J$ they extend to about $5$ lattice sites.

\section{Summary}
\label{sec.sum}

Using the Bogoliubov--de~Gennes technique we have performed numerical calculations to obtain the eigenvalues and eigenvectors of a system comprised of two magnetic impurities in a superconductor with triangular lattice. 
We have shown that the presence of a dimer induces two quantum phase transitions, each of which increases the polarization of the system by $1/2$. 
Upon calculating the local density of states we have found the in-gap spectrum of hybridized YSR bound states, and their spatial character. 
The topography of the bound states exhibits a breaking of rotational symmetry, due to the overlap of YSR wavefunctions coming from double impurities. 
The shape presented in the LDOS maps changes with the value of magnetic interaction $J$, albeit the spatial extent of bound states is not extended for higher values of coupling.

\begin{acknowledgments}
We thank Aksel Kobia\l{}ka and Tadeusz Doma\'{n}ski for valuable comments and discussions.
This work was supported by the National Science Centre (NCN, Poland) grant UMO-2017/25/B/ST3/02586  (A.P.).
\end{acknowledgments}

\bibliography{biblio}

\end{document}